\documentclass{article}

\usepackage[final]{neurips_2021}

\usepackage[utf8]{inputenc} 
\usepackage[T1]{fontenc}    
\usepackage{graphicx}       
\usepackage{hyperref}       
\usepackage{booktabs}       
\usepackage{amsfonts}       
\usepackage{nicefrac}       
\usepackage{microtype}      
\usepackage{xcolor}         
\usepackage{lscape}         

\title{Synthetic weather radar using hybrid quantum-classical machine learning}

\author{
    Graham R. Enos\\
    Rigetti Computing\\
    \texttt{genos@rigetti.com}\\    
    \And
    Matthew J. Reagor\\
    Rigetti Computing\\
    \texttt{matt@rigetti.com}\\
    \And
    Maxwell P Henderson\\
    Rigetti Computing\\
    \And
    Christina Young\\
    Rigetti Computing\\
    \And
    Kyle Horton\\
    Rigetti Computing\\
    \And
    Mandy Birch\\
    Rigetti Computing\\
    \And
    Chad Rigetti\\
    Rigetti Computing\\
}

\begin{document}

\maketitle

\begin{abstract}
    The availability of high-resolution weather radar images underpins effective forecasting and decision-making.
    In regions beyond traditional radar coverage, generative models have emerged as an important synthetic capability, fusing more ubiquitous data sources, such as satellite imagery and numerical weather models, into accurate radar-like products.
    Here, we demonstrate methods to augment conventional convolutional neural networks with quantum-assisted models for generative tasks in global synthetic weather radar.
    We show that quantum kernels can, in principle, perform fundamentally more complex tasks than classical learning machines on the relevant underlying data.
    Our results establish synthetic weather radar as an effective heuristic benchmark for quantum computing capabilities and set the stage for detailed quantum advantage benchmarking on a high-impact operationally relevant problem.
\end{abstract}

\section{Introduction}
Global Synthetic Weather Radar (GSWR) is a class of techniques for assimilating diverse meteorological data types, in order to produce synthetic weather radar images. An archetype use-case for GSWR is air traffic management for flights in remote regions, beyond the reach of high-resolution radar coverage. A leading GSWR model was developed for production use by the team at MIT LL, known as the Offshore Precipitation Capability (OPC) presented in~\citet{veillette2018creating}. The OPC-CNN is a machine learning (ML) model based on convolutional neural networks (CNN's) that integrates several kinds of high dimensional weather data at different spatial scales and temporal resolutions. The performance of OPC-CNN has already driven its adoption for real-world operations. Yet, challenges remain to improve its reliability, relative to true radar infrastructure.

In recent years, there has been tremendous effort to understand the role of quantum information processing to improve ML tasks, based on either quantum-assisted training of classical models, where quadratic or polynomial speed-ups are anticipated, or with data encoded into qubits, where exponential advantage is a possibility~\citep{huang2021power}. While various candidate data types have been explored, recently, a geometric metric over such data was proposed in~\citet{huang2021power} to determine the suitability of a problem domain to quantum machine learning. An open question has been if real-world data has sufficient complexity to satisfy this metric, and, if so, whether a viable model could be constructed and benchmarked for that data through quantum ML methods.

In this work, we provide evidence that the input data to GSWR problems, specifically those in the OPC-CNN system, can have a structure theoretically compatible with quantum advantage for some kernel functions. Next, we develop two case studies to investigate the compatibility of OPC-CNN with hybrid quantum-classical CNN's. First, we construct an end-to-end system based on OPC-CNN, including model calibration and evaluation, that replaces one of the data input streams with synthetic data, collected from a pre-trained quantum ML model. This hybrid system performs as well as the baseline OPC-CNN system with access to ground-truth input data. As a second case study, we evaluate replacing convolutional layers in the OPC-CNN model with quantum convolutional (quanvolutional) layers and observe competitive model performance, despite the small, noisy quantum hardware under test. Finally, we comment on next steps towards developing quantum acceleration for GSWR.

\section{Application Context}    
\begin{figure}
    \centering
    \includegraphics[width=0.7\textwidth]{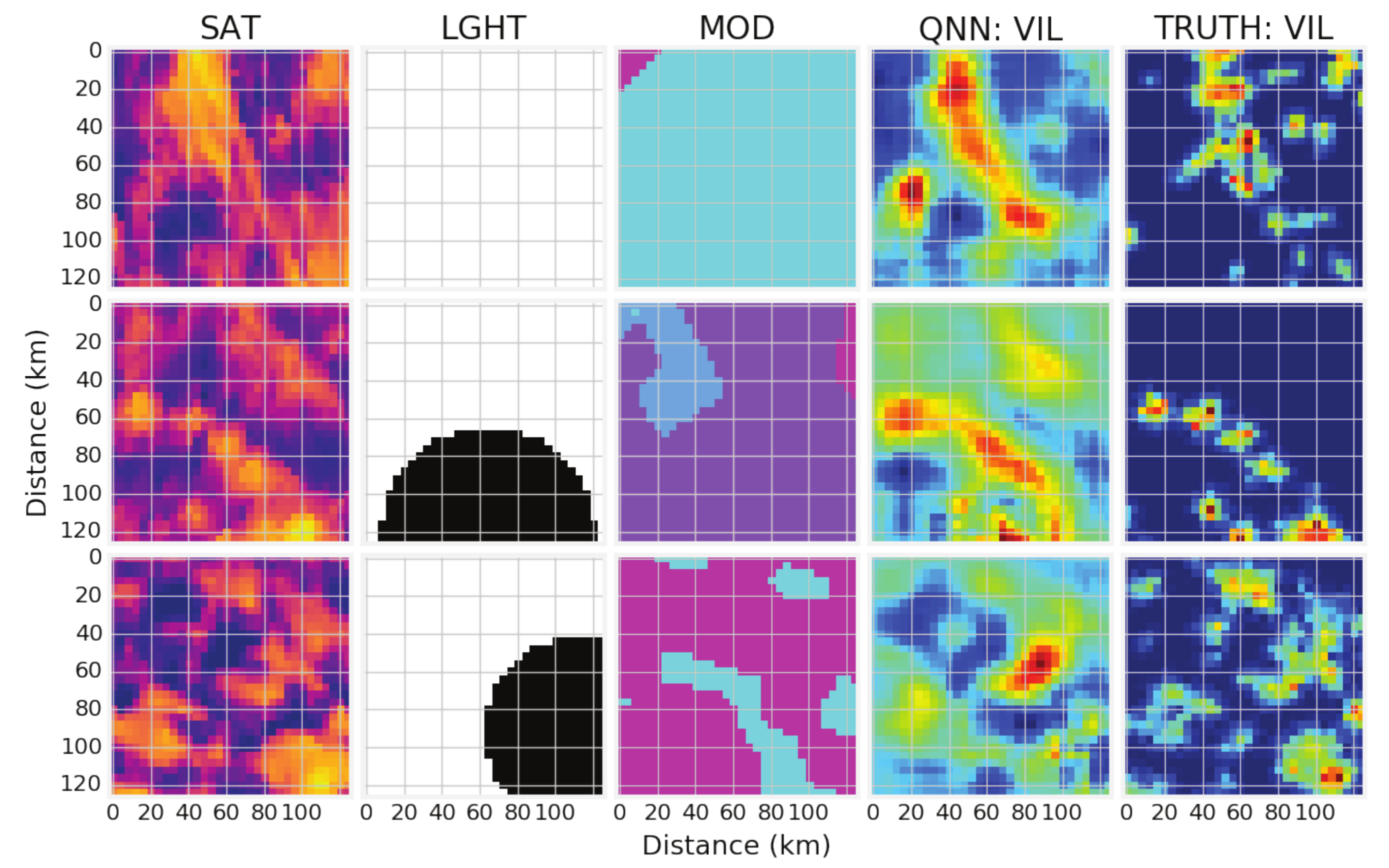}
    \caption{Three examples from the test set are organized into rows, showing a subset of the input data  and a subset of weather products from the hardware-trained QNN model and the ground truth measurements. All patches are 128 km \(\times\) 128 km at a 4 km resolution. Visual inspection with the true radar shows good agreement, with false alarms in the QNN corresponding to an over-reliance on SAT features. We observe limitations on predicting fine-features (few km scale), which is expected for the small circuit size used for this proof of concept. 
}
    \label{fig:qnn_vil}
\end{figure}

The overall goal for GSWR is to infer weather radar imagery from input data, which comprises, for OPC-CNN (see Fig~\ref{fig:qnn_vil}): satellite imagery (SAT), lighting strike observations (LGHT), and numerical weather models (MOD). As output, we seek three types of common weather radar products: vertical integrated liquid (VIL), echo tops (ET), and composite reflectivity (CR), which can be used to guide operational planning. A brief description of these input and output data types is summarized in Table~\ref{tab:data}; further details can be found in~\citet{veillette2018creating} and~\citet{roebber2009visualizing}. Importantly, GSWR models can be used to infer weather conditions when traditional radar products are unavailable, such as in remote regions, or in disaster response. Training data is taken from sectors with true radar coverage.

Establishing a set of performance criteria for inferred weather conditions is critical to mission success and practical model training.
Fortunately, the meteorological community has a long history of tracking forecast quality \citep{stanski1989survey, schaefer1990critical, roebber2009visualizing}, and these metrics can be used to benchmark synthetic data as per~\citet{veillette2018creating}.
The classification of individual pixels breaks down into four categories: false alarms, hits, misses, and correct rejections.
Based on these pixel-level values, model evaluation consists of four key statistics~\citep{roebber2009visualizing}: (1) the bias statistic (BIAS) which estimates whether a system predicts more inclement weather than ground truth overall; (2) the probability of detection (POD) which estimates the fraction of pixels correctly predicted as events relative to the total pixel count of inclement weather; (3) the success rate (SUCR) which is one minus the false alarm rate, the fraction of false alarm pixels; (4) the critical success index (CSI) which is the probability of a true detection after accounting for false alarms. We report these metrics for the case studies that follow.

\begin{table*}[ht]
    \caption{Data sets for synthetic weather radar.}
    \label{tab:data}
    \centering
\begin{tabular}{p{1cm}p{6cm}p{1cm}p{1cm}p{1cm}p{1cm}}
\toprule
Channel & Description & Pixels & Layers & Res. & Type \\
\midrule
SAT & Cloud top height, solar zenith angle, visible band (600nm), and 4 IR bands & 32x32 & 7 & 4~km & Input \\
LGHT & Gaussian-blurred lightning strike location histories in 10min, 20min, and 30min  & 32x32 & 3 & 4~km & Input \\
MOD & Impacting fields from numerical weather models including temperature, pressure  & 32x32 & 7 & 4~km & Input \\
TARG & Target products: vertical integrated liquid, echo-top, and composite reflectivity  & 32x32 & 3 & 4~km & Output \\
\bottomrule
\end{tabular}
\end{table*}

\section{Results}
\begin{figure}
  \centering
  \includegraphics[width=0.6\textwidth]{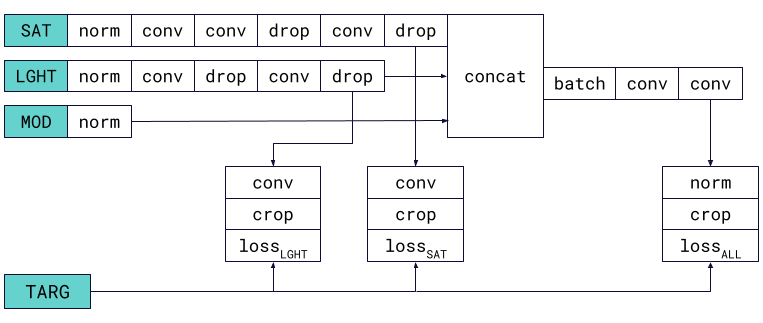}
  \caption{OPC-CNN architecture.}
  \label{fig:arch}
\end{figure}

The OPC-CNN architecture is shown in Figure~\ref{fig:arch}.
Functionally, this system is trained to extract key features from the input data, to combine those features, and to make useful inferences towards the three GSWR weather products based on that combined feature-set.
The model---implemented in \texttt{TensorFlow}~\citep{tensorflow}---first passes input data in three different modalities through a series of feature extraction layers.
These three extraction pipelines are trained against the target output data prior to passing through the fusion layer which is again trained against the target.

\subsection{Geometric Difference Metric}

In \citet{huang2021power}, the authors outline a protocol to compare how two kernel methods, one classical and one quantum, would perform on a specified data set.
This protocol involves computing a kernel matrix for each method and then looking at the geometric difference between the two.
If the difference is favorably sized with respect to the number of data points involved, then a complexity test is performed for a specific classification label.
Should the complexity test yield a favorable result, there is a potential for significant quantum advantage for the problem instance.

We applied these tests to the OPC-CNN model. 
After restricting the data to the first \(M\) principal components for \(M\) in \(\{4, 8, 16, 32\}\), we computed the classical kernel matrix \(K_C = DD^\mathsf{T}\) from the \(N \times M\) data matrix \(D\).
Computing the quantum kernel depends on a data encoding scheme.
The first was a simple angle encoding, with the values (after appropriate scaling) used as rotation angles in \texttt{RX} quantum gates.
The second was a more complex instantaneous quantum polynomial (IQP)-style circuit as in \citet{havlivcek2019supervised} and \citet{huang2021power}.
Given two feature vectors \(x_i\) and \(x_j\) (rows of \(D\)) and the encoding \(E\), we executed the quantum circuit \(E(x_i)E^\dagger(x_j)\) and counted the number of all-zero bitstrings occurring in the returned samples to empirically estimate the value \(\left\vert \langle 0 \vert  E(x_i)E^\dagger(x_j) \vert 0 \rangle \right\vert ^2\), filling in the \(i,j\) and \(j,i\) entries of the quantum kernel matrix \(K_Q\).
We then computed the geometric difference
\(g(K_c \Vert K_Q) = \sqrt{ \left\Vert \sqrt{K_Q} K_C^{-1} \sqrt{K_Q} \right\Vert_\infty }\).
If this difference \(g\) is close to the square root of \(N\), there exists a labelling for this data set such that the quantum classifier will likely outperform the classical one per~\citet{huang2021power}.

We sampled \(N = 74\) feature vectors for simulated noiseless processor sizes (4, 8, and 16 qubits) and for 32 physical qubits on the Rigetti Aspen-9 processor.
The geometric difference at 32 qubits for both data sources and encoding schemes were close to \(N\), indicating that a labelling exists for which a quantum classifier would likely outperform the classical one.
At smaller, QVM-simulated qubit sizes, the geometric differences were similarly favorable.
Though \citet{huang2021power} guarantees, therefore, that a label exists for which a quantum classifier would be expected to outperform a classical one, that label is not necessarily related to the TARG variable. 

For a specific labelling or target variable, then, \citet{huang2021power} proposes a secondary test comparing the complexities \(s_C\) and \(s_Q\) of support vector machine classifiers (from \texttt{scikit-learn}~\citep{scikit-learn}) trained on the two kernel matrices by taking the 2-norm (i.e. the largest singular value) of their dual coefficients. Given the favorable \(g\) values, we computed the \(s_C\) and \(s_Q\) at all sizes and encodings for the labels given by the target values from the synthetic weather radar data set.
For each of the two data encodings on all qubit sizes, both simulated and full 32 qubits on hardware, \(s_Q\) was larger than the classical matrix’s \(s_C\), and additionally, that \(s_C\) was smaller than \(\sqrt{N}\), indicating that the classical classifiers predicted TARG better than the quantum classifiers at these sizes. 
The nature of the theoretical quantum advantage associated with the GSWR data remains an outstanding question.

\subsection{Quantum Variational Autoencoder}

As a first case study towards developing heuristic tests, we developed a generative quantum model to mimic one of the two main data sources in the event of data scarcity or unreliability.
Both of these sources can prove unreliable at times, and a model trained to produce data resembling the historical distribution of the original source could fill that strategic gap.
For a given data source (LGHT or SAT), we constructed generative models as follows.
First, we trained a vector-quantized variational autoencoder (VQVAE), a modified VAE which restricts the latent embedding space to a discrete set of “codebook” vectors, on the data source.
During training, each input vector in the latent space moves to the nearest codebook entry~\citep{van2017neural}.
Once VQVAE training was complete, we encoded the training data and converted it to bitstrings to train a quantum circuit Born machine (QCBM), a generative quantum model that produces bitstrings from the empirical distribution of a given collection \citep{coyle2021quantum}.
For the best performance, a QCBM was first trained on the QVM, then the QVM’s best parameters were used as the starting point for training a QCBM on the QPU.
The \texttt{Faiss} library \citep{johnson2017billion} was used to match samples with the closest codebook bitstring via nearest neighbor lookup to mitigate errors.
Next, we created the full generative model by sending QCBM-sampled data through the VQVAE’s decoder.
Enough samples were gathered from this quantum VAE to replace the corresponding true data source and then train the full OPC-CNN.
These experiments were run with 16 qubits, each corresponding to a single entry in the VQVAE’s latent codebook.

We find that the VQVAE was more effective with the sparser lightning data than the more dense and complex satellite data.
The lightning model’s test metrics were on par with the classical at lower elevation levels, though there is some taper at higher ones, suggesting that the model generates synthetic lightning data better for lower level storms.
This demonstrates the promise of this methodology and the need for refinement and parameter tuning for stronger operational applicability.
The generative models enabled the utilization of the full OPC-CNN training and validation setup, including the model calibration step.
Per \citet{veillette2018creating}, this calibration addresses a model’s BIAS, adjusting it towards the ideal value of one with a histogram matching procedure.
Furthermore, validation metrics can be computed over the full test images.
In the left portion of Figure~\ref{fig:sucr_pod_vil}, the effect of model calibration is shown, with values consistently pulled towards the diagonal line where POD = SUCR.
In the same figure, we can see how the full validation apparatus of OPC-CNN enabled the examination of the models’ performance at various thresholds, both with and without calibration.
This plot shows the product points at various thresholds against the contours of CSI.
As most points for the LGHT generative model are close to the POD = SUCR diagonal and are distributed similarly to the classical model with respect to the contours of the critical success index, it shows success in simulating missing lightning data.

\subsection{Quanvolutional Neural Network}

A second case study leveraged quantum-convolutional (quanvolutional) neural networks, improving on some metrics while requiring modifications to the software architecture.
For each of the two non-MOD data sources, LGHT and SAT, we replaced the first normalization and convolutional blocks with a randomized quanvolutional block, while the rest of the network and data remained unchanged.

A sampling-based approach was employed due to the large amount of training data (almost 75,000 examples, each consisting of either 7,168 or 3,072 data points, which required excessive bandwidth for a small, noisy quantum processor in a reasonable amount of time).
We trained the quanvolutional layer in randomized order for three and a half hours of processing time on the Aspen-9 processor, with the input/output pairs saved. Once the quantum processing completed, a similarity search was performed using the \texttt{Faiss} library \citep{johnson2017billion} to perform a nearest-neighbor lookup, following insight from \citet{henderson2021methods}.
Given new input data, we found the nearest centroid and returned its paired output.
This approach exhibited improved performance over \citet{henderson2021methods} due to \texttt{Faiss} being graphic processing unit (GPU) enabled.
As in \citet{henderson2020quanvolutional}, this setup attempts to learn as much or more as a classical CNN block but with fewer parameters covering a larger available feature space.
As before, two different encoding schemes were evaluated for loading classical data into the quantum processor: an angle encoding, and an IQP-style encoding~\citep{havlivcek2019supervised, huang2021power}.

The QNN models for the lightning data were successful; the angle-encoded lightning model outperformed the classical model at level 2, improving VIL CSI from 0.47 (both uncalibrated and calibrated classical model) to 0.48 and VIL BIAS from 0.83 or 0.85 (uncalibrated and calibrated, respectively) to 0.92 (recall that 1 is ideal and the QNN undergoes no post-processing calibration), as shown on the right of Figure~\ref{fig:sucr_pod_vil}.
At level 2, the same QNN model improved on other metrics as well; see the appendix for the complete set of metrics.
It should be reiterated, though, that these improvements occurred at level 2; storms of higher severity are harder to predict.
For the satellite data, classical models outperformed four different hybrid QNN models (two different data sources quanvolved with two different encoding schemes), using two key metrics, CSI and BIAS.
While the quanvolutional setup requires modifying the OPC-CNN architecture and thus cannot undergo calibration without also quanvolving the calibration data, the best performing QNN model surpassed the calibrated classical model in key metrics, including BIAS which calibration is intended to improve. 

\begin{figure}
  \centering
  \includegraphics[width=0.6\textwidth]{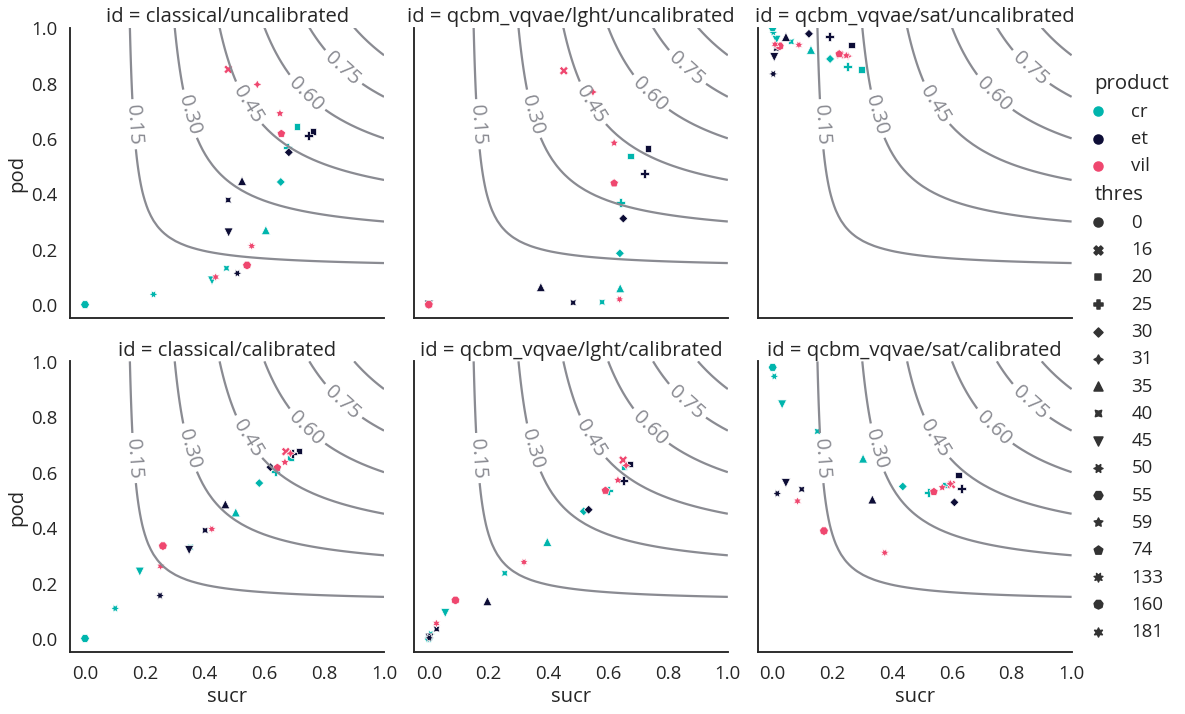}
  \includegraphics[width=0.35\textwidth]{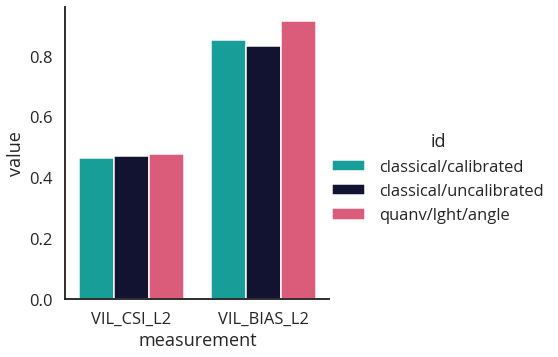}
  \caption{Comparing metrics of Quantum VAEs and classical models.}
  \label{fig:sucr_pod_vil}
\end{figure}

\section{Conclusion and Future Work}

These results are initial evidence that data in real-world ML problems, here high dimensional weather data, can have a structure theoretically compatible with quantum advantage.
Based on those findings, we developed two case studies that demonstrate how to hybridize a start of the art GSWR system with quantum ML techniques.
Both models showed promise with respect to operationally relevant meteorological performance metrics.
Ongoing development of the methods presented, alongside anticipated improvements in quantum computing system performance, indicate substantial promise for the role of quantum processing in GSWR and related problems.

This research was, in part, funded by the U.S. Government. The views and conclusions contained in this document are those of the authors and should not be interpreted as representing the official policies, either expressed or implied, of the U.S. Government.

\bibliographystyle{unsrtnat}
\bibliography{main}

\begin{thebibliography}{13}
\providecommand{\natexlab}[1]{#1}
\providecommand{\url}[1]{\texttt{#1}}
\expandafter\ifx\csname urlstyle\endcsname\relax
  \providecommand{\doi}[1]{doi: #1}\else
  \providecommand{\doi}{doi: \begingroup \urlstyle{rm}\Url}\fi

\bibitem[Veillette et~al.(2018)Veillette, Hassey, Mattioli, Iskenderian, and
  Lamey]{veillette2018creating}
Mark~S. Veillette, Eric~P. Hassey, Christopher~J. Mattioli, Haig Iskenderian,
  and Patrick~M. Lamey.
\newblock Creating synthetic radar imagery using convolutional neural networks.
\newblock \emph{Journal of Atmospheric and Oceanic Technology}, 35\penalty0
  (12):\penalty0 2323--2338, 2018.
\newblock \doi{10.1175/JTECH-D-18-0010.1}.

\bibitem[Huang et~al.(2021)Huang, Broughton, Mohseni, Babbush, Boixo, Neven,
  and McClean]{huang2021power}
Hsin-Yuan Huang, Michael Broughton, Masoud Mohseni, Ryan Babbush, Sergio Boixo,
  Hartmut Neven, and Jarrod~R McClean.
\newblock Power of data in quantum machine learning.
\newblock \emph{Nature communications}, 12\penalty0 (1):\penalty0 1--9, 2021.

\bibitem[Roebber(2009)]{roebber2009visualizing}
Paul~J Roebber.
\newblock Visualizing multiple measures of forecast quality.
\newblock \emph{Weather and Forecasting}, 24\penalty0 (2):\penalty0 601--608,
  2009.

\bibitem[Stanski et~al.(1989)Stanski, Wilson, and Burrows]{stanski1989survey}
Henry~R Stanski, Laurence~J Wilson, and William~R Burrows.
\newblock Survey of common verification methods in meteorology.
\newblock \emph{World Weather Watch Technical Report}, 1989.

\bibitem[Schaefer(1990)]{schaefer1990critical}
Joseph~T Schaefer.
\newblock The critical success index as an indicator of warning skill.
\newblock \emph{Weather and forecasting}, 5\penalty0 (4):\penalty0 570--575,
  1990.

\bibitem[Developers(2021)]{tensorflow}
TensorFlow Developers.
\newblock Tensorflow, August 2021.
\newblock URL \url{https://doi.org/10.5281/zenodo.5189249}.

\bibitem[Havl{\'\i}{\v{c}}ek et~al.(2019)Havl{\'\i}{\v{c}}ek, C{\'o}rcoles,
  Temme, Harrow, Kandala, Chow, and Gambetta]{havlivcek2019supervised}
Vojt{\v{e}}ch Havl{\'\i}{\v{c}}ek, Antonio~D C{\'o}rcoles, Kristan Temme,
  Aram~W Harrow, Abhinav Kandala, Jerry~M Chow, and Jay~M Gambetta.
\newblock Supervised learning with quantum-enhanced feature spaces.
\newblock \emph{Nature}, 567\penalty0 (7747):\penalty0 209--212, 2019.

\bibitem[Pedregosa et~al.(2011)Pedregosa, Varoquaux, Gramfort, Michel, Thirion,
  Grisel, Blondel, Prettenhofer, Weiss, Dubourg, Vanderplas, Passos,
  Cournapeau, Brucher, Perrot, and Duchesnay]{scikit-learn}
F.~Pedregosa, G.~Varoquaux, A.~Gramfort, V.~Michel, B.~Thirion, O.~Grisel,
  M.~Blondel, P.~Prettenhofer, R.~Weiss, V.~Dubourg, J.~Vanderplas, A.~Passos,
  D.~Cournapeau, M.~Brucher, M.~Perrot, and E.~Duchesnay.
\newblock Scikit-learn: Machine learning in {P}ython.
\newblock \emph{Journal of Machine Learning Research}, 12:\penalty0 2825--2830,
  2011.

\bibitem[van~den Oord et~al.(2017)van~den Oord, Vinyals, and
  Kavukcuoglu]{van2017neural}
Aaron van~den Oord, Oriol Vinyals, and Koray Kavukcuoglu.
\newblock Neural discrete representation learning.
\newblock In \emph{Proceedings of the 31st International Conference on Neural
  Information Processing Systems}, pages 6309--6318, 2017.

\bibitem[Coyle et~al.(2021)Coyle, Henderson, Le, Kumar, Paini, and
  Kashefi]{coyle2021quantum}
Brian Coyle, Maxwell Henderson, Justin Chan~Jin Le, Niraj Kumar, Marco Paini,
  and Elham Kashefi.
\newblock Quantum versus classical generative modelling in finance.
\newblock \emph{Quantum Science and Technology}, 6\penalty0 (2):\penalty0
  024013, 2021.

\bibitem[Johnson et~al.(2017)Johnson, Douze, and J{\'e}gou]{johnson2017billion}
Jeff Johnson, Matthijs Douze, and Herv{\'e} J{\'e}gou.
\newblock Billion-scale similarity search with gpus.
\newblock \emph{arXiv preprint arXiv:1702.08734}, 2017.

\bibitem[Henderson et~al.(2021)Henderson, Gallina, and
  Brett]{henderson2021methods}
Max Henderson, Jarred Gallina, and Michael Brett.
\newblock Methods for accelerating geospatial data processing using quantum
  computers.
\newblock \emph{Quantum Machine Intelligence}, 3\penalty0 (1):\penalty0 1--9,
  2021.

\bibitem[Henderson et~al.(2020)Henderson, Shakya, Pradhan, and
  Cook]{henderson2020quanvolutional}
Maxwell Henderson, Samriddhi Shakya, Shashindra Pradhan, and Tristan Cook.
\newblock Quanvolutional neural networks: powering image recognition with
  quantum circuits.
\newblock \emph{Quantum Machine Intelligence}, 2\penalty0 (1):\penalty0 1--9,
  2020.

\end{thebibliography}

\pagebreak

\appendix
\section{Geometric Difference and Secondary Complexity}

\begin{figure}
    \centering
    \includegraphics[width=0.33\textwidth]{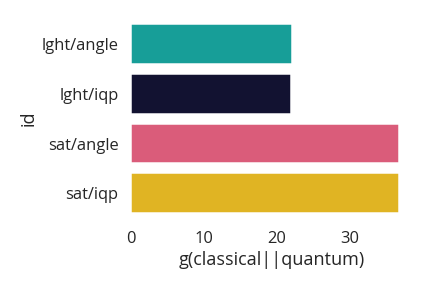}
    \includegraphics[width=0.5\textwidth]{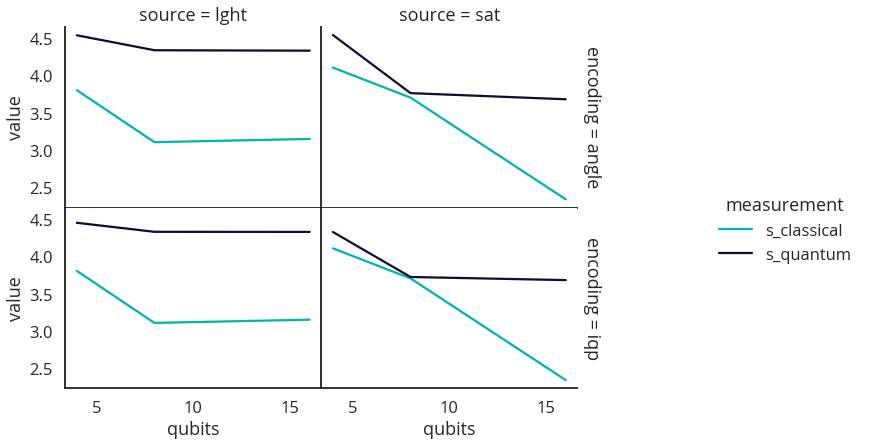}
    \caption{Evidence for theoretical quantum advantage. (left) Geometric difference favored quantum advantage, (right) secondary tests indicate the standard labels of OPC-CNN lack sufficient complexity to generate that advantage.}
    \label{fig:geo_diff}
\end{figure}

We studied the geometric differences and secondary complexities of four different combinations of data source (LGHT or SAT) and encoding schemes (angle or IQP).
On the left of Figure~\ref{fig:geo_diff}, we can see that the values of \(g(K_C \Vert K_Q)\) were all larger than \(\sqrt{N} = \sqrt{74}\) for the \(K_Q\) computed on the QPU.
However, none of the ratios of the secondary complexities \(s_C\) and \(s_Q\) favored the quantum kernel; as the right portion of Figure~\ref{fig:geo_diff} shows, the classical complexity was lower at all simulated qubit sizes.

\section{Test Metrics}

Table~\ref{tab:full_metrics} contains the available test metrics (mean squared error, critical success index, bias, and probability of detection) of the three products (vertical integrated liquid, composite reflectivity, and echo-top) for each of the models studied.

\begin{landscape}
\begin{table}
    \caption{Metrics for all models.}
    \label{tab:full_metrics}
    \centering
\begin{tabular}{llrrrrrrrrrrr}
\toprule
           model & VIL MSE &  ET MSE &  CR MSE &  VIL CSI &  ET CSI &  CR CSI &  VIL BIAS &  ET BIAS &  CR BIAS &  VIL POD &  ET POD &  CR POD \\
\midrule
   classical cal &   510.7 &   24.57 &   51.00 &     0.47 &    0.53 &    0.45 &      0.85 &     0.79 &     0.74 &     0.59 &    0.62 &    0.54 \\
 classical uncal &   492.0 &   23.43 &   49.38 &     0.47 &    0.53 &    0.45 &      0.83 &     0.77 &     0.72 &     0.59 &    0.61 &    0.53 \\
   qvae lght cal &   578.7 &   28.84 &   55.04 &     0.44 &    0.48 &    0.43 &      0.88 &     0.87 &     0.74 &     0.57 &    0.61 &    0.52 \\
 qvae lght uncal &   568.7 &   25.92 &   55.31 &     0.39 &    0.45 &    0.34 &      0.67 &     0.69 &     0.52 &     0.46 &    0.53 &    0.38 \\
    qvae sat cal & 2,671.3 &  179.16 &  343.74 &     0.37 &    0.32 &    0.41 &      2.20 &     2.90 &     1.91 &     0.87 &    0.95 &    0.85 \\
  qvae sat uncal & 5,021.3 &  303.42 &  582.92 &     0.35 &    0.28 &    0.40 &      2.51 &     3.44 &     2.11 &     0.91 &    0.97 &    0.89 \\
quanv lght angle &   524.1 &   24.87 &   51.20 &     0.48 &    0.54 &    0.47 &      0.92 &     0.84 &     0.80 &     0.62 &    0.64 &    0.57 \\
  quanv lght iqp &   513.4 &   24.12 &   50.30 &     0.47 &    0.52 &    0.45 &      0.86 &     0.79 &     0.74 &     0.59 &    0.62 &    0.54 \\
 quanv sat angle &   677.9 &   34.51 &   70.40 &     0.30 &    0.37 &    0.26 &      0.48 &     0.48 &     0.39 &     0.34 &    0.40 &    0.28 \\
   quanv sat iqp &   887.0 &   52.94 &   94.15 &     0.21 &    0.23 &    0.16 &      0.31 &     0.32 &     0.23 &     0.22 &    0.25 &    0.17 \\
\bottomrule
\end{tabular}
\end{table}
\end{landscape}

\end{document}